\documentclass[reprint,amsmath,amssymb,aps,superscriptaddress,nofootinbib,floatfix,prl,preprintnumbers]{revtex4-1}
\usepackage{graphicx}
\usepackage{dcolumn}
\usepackage{bm}
\usepackage{hyperref}
\usepackage{xcolor}
\usepackage{nicefrac}
\usepackage{soulpos}
\ulposdef{\hlst}{%
\rlap{\textcolor{yellow}{\rule[-.75ex]{\ulwidth}{2.5ex}}}%
\rule[.45ex]{\ulwidth}{.1ex}%
}

\usepackage{hyperref}
\hypersetup{
    colorlinks=true,
    urlcolor=blue,
    linkcolor = blue,
    urlcolor  = blue,
    citecolor = blue,
    anchorcolor = blue,
    pdftitle={...},
    pdfauthor={...},
    pdfsubject={...}
    }
\newcommand{\GeV}{{\rm GeV}}

\begin{document}

\title{
Global Data-Driven Determination of Baryon Transition Form Factors
}

\author{Yu-Fei Wang}
\email{wangyufei@ucas.ac.cn}
\affiliation{Institute for Advanced Simulation (IAS-4), 
Forschungszentrum J\"ulich, 52425 J\"ulich, Germany}
\affiliation{School of Nuclear Science and Technology, University of Chinese Academy of Sciences, Beijing 101408, China}

\author{Michael D{\"o}ring}
\email{doring@gwu.edu}
\affiliation{Institute for Nuclear Studies and Department of Physics, 
The George Washington University, Washington, DC 20052, USA}
\affiliation{Thomas Jefferson National Accelerator Facility, 
Newport News, Virginia 23606, USA}

\author{Jackson Hergenrather}
\affiliation{Institute for Nuclear Studies and Department of Physics, 
The George Washington University, Washington, DC 20052, USA}

\author{Maxim Mai}
\email{mai@hiskp.uni-bonn.de}
\affiliation{Helmholtz-Institut f\"ur Strahlen- und Kernphysik (Theorie) and Bethe Center for Theoretical Physics, 
Universit\"at Bonn, 53115 Bonn, Germany}
\affiliation{Institute for Nuclear Studies and Department of Physics, 
The George Washington University, Washington, DC 20052, USA}

\author{Terry Mart}
\email{terry.mart@sci.ui.ac.id}
\affiliation{Departemen Fisika, FMIPA, Universitas Indonesia, Depok 16424, Indonesia}

\author{Ulf-G.~Mei\ss ner} \email{meissner@hiskp.uni-bonn.de}
\affiliation{Helmholtz-Institut f\"ur Strahlen- und Kernphysik (Theorie) and Bethe Center for Theoretical Physics, 
Universit\"at Bonn, 53115 Bonn, Germany}
\affiliation{Institute for Advanced Simulation (IAS-4), 
Forschungszentrum J\"ulich, 52425 J\"ulich, Germany}
\affiliation{Tbilisi State University, 0186 Tbilisi, Georgia}

\author{Deborah R{\"o}nchen}
\email{d.roenchen@fz-juelich.de}
\affiliation{Institute for Advanced Simulation (IAS-4), 
Forschungszentrum J\"ulich, 52425 J\"ulich, Germany}

\author{Ronald Workman}
\affiliation{Institute for Nuclear Studies and Department of Physics, 
The George Washington University, Washington, DC 20052, USA}

\collaboration{J\"ulich-Bonn-Washington Collaboration}

\begin{abstract}
Hadronic resonances emerge from strong interactions encoding the dynamics of quarks and gluons. The structure of these resonances can be probed by virtual photons parametrized in transition form factors. In this study, twelve $N^*$ and $\Delta$ transition form factors at the pole are extracted from data with the center-of-mass energy from $\pi N$ threshold to $1.8\,\GeV$, and the photon virtuality $0\leq Q^2/\GeV^2\leq 8$. For the first time, these results are determined from a simultaneous analysis of more than one state, i.e., $\sim 10^5$ $\pi N$, $\eta N$, and $K\Lambda$ electroproduction data. In addition, about $ 5\cdot 10^4$ data in the hadronic sector as well as photoproduction serve as boundary conditions. For the $\Delta(1232)$ and $N(1440)$ states our results are in qualitative agreement with previous studies, while the transition form factors at the poles of some  higher excited states are estimated for the first time. Realistic uncertainties are determined by further exploring the parameter space.
\end{abstract}

\preprint{JLAB-THY-24-4016}

\maketitle

{\it Introduction}--The spectrum and structure of hadronic resonances encode the dynamics of Quantum Chromodynamics (QCD) at intermediate energies where confinement and chiral symmetry breaking lead to the properties of matter as we know it. See Ref.~\cite{Mai:2022eur} for a recent review.  Especially the excited states of the nucleon ($N^*$) and $\Delta$ resonances, have been studied for decades. On the experimental side, electromagnetic interactions are clean probes of the structure of hadrons, motivating the study of resonances through meson photoproduction reactions~\cite{Ireland:2019uwn,Thiel:2022xtb}. Electroproduction processes~\cite{Aznauryan:2012ba, Mokeev:2022xfo, Carman:2020qmb, Blin:2021twt, HillerBlin:2022ltm} provide additional information with an extra degree of freedom, the photon virtuality $Q^2$, serving as an energy scale to probe hadron structure, particularly through the electromagnetic transition form factors (TFFs) between excited and ground state baryons~\cite{Aznauryan:2011qj, Ramalho:2023hqd} facilitated by experimental progress~\cite{CLAS:2002xbv, CLAS:2009ces, CLAS:2012wxw,  Mokeev:2015lda}. The structure encoded in TFFs allows conclusions on the nature of resonances like the $N(1440)$~\cite{Burkert:2017djo}; they also determine quark transverse charge densities~\cite{Carlson:2007xd, Tiator:2009mt, Miller:2010nz}, and even provide critical information for the exploration of possible hybrid baryons with the gluon as a constituent~\cite{JLABhybrid}.

On the theory side, quark models and Dyson-Schwinger approaches~\cite{Eichmann:2016yit,Merten:2002nz, Obukhovsky:2011sc,Eichmann:2011aa,Wilson:2011aa,Segovia:2014aza,Segovia:2015hra,Aznauryan:2018okk,Obukhovsky:2019xrs,Burkert:2017djo,Chen:2018nsg,Chen:2023zhh} assume the relevant $N^*$ and $\Delta$ states are three-quark cores plus possible configurations at the hadron level. In the small-$Q^2$ region, the virtual photon mainly interacts with the peripheral components like the meson cloud~\cite{Ramalho:2017iqq}, while when $Q^2$ is larger the properties of the three-quark core shine through. However, there is no unique separation of these regimes. Calculations of electromagnetic form factors in lattice QCD are advancing, as well~\cite{Lin:2008qv, Aliev:2013dxa,Agadjanov:2014kha, Lin:2011ti,Radhakrishnan:2022ubg,Stokes:2024haa,Liu:2020jmn}. Chiral Perturbation Theory approaches and extensions thereof also allow for the calculation of multipoles and resonance TFFs~\cite{Bernard:1996bi, Gail:2005gz, Jido:2007sm, Doring:2010rd, Hilt:2017iup,Ruic:2011wf,Mai:2012wy}, they are, however, restricted to individual resonances and reaction channels in a limited energy region. For a recent extraction of $\Delta$ and $N(1440)$ pole parameters based on Roy-Steiner equations, see~\cite{Hoferichter:2023mgy}.

Transition form factors should be independent of the reaction through which they are determined, calling for the simultaneous analysis of multiple final states, as carried out for the first time in this research effort. This is achieved by employing the J{\"u}lich-Bonn-Washington (JBW) approach, a dynamical coupled-channel model for the study of pion-, photon- and electron-induced reactions. Simultaneously analyzing different final states also facilitates the difficult access to the TFFs of less prominent baryon resonances. For example, we compare here to the TFFs of the $\Delta(1600)$ determined recently by analyzing the $\pi\pi N$ final state~\cite{Mokeev:2023zhq} complementing theory predictions~\cite{Lu:2019bjs}. 

Indeed, truly reaction-independent TFFs can only be defined at resonance poles. In the current study, this is achieved by straightforward analytic continuation of multipoles to  complex energies, similar to what was achieved in another dynamical coupled-channel approach for some resonances, the ANL-Osaka model~\cite{Julia-Diaz:2006ios,Suzuki:2010yn,Kamano:2018sfb}. TFFs in unitary isobar models~\cite{Drechsel:1998hk,Tiator:2003uu,Drechsel:2007if,Tiator:2011pw,Hilt:2013fda} are usually reported in terms of Breit-Wigner parameters~\cite{Tiator:2011pw}, but secondary parametrizations and fits can be used to access TFFs at the pole, as well~\cite{Workman:2013rca, Tiator:2016btt}. 

Electroproduction data are incomplete resulting in ambiguities~\cite{Tiator:2017cde, JointPhysicsAnalysisCenter:2023gku} in extracted multipoles that bear the TFFs. Indeed, polarization measurements in electroproduction are sparse~\cite{JeffersonLabHallA:2005wfu, CLAS:2022yzd}. A strategy to represent the sparsity and incompleteness of data is to explore the parameter space and pin down multiple local minima in the $\chi^2$ optimization. As these uncertainties dominate all others, the multipoles and TFFs extracted in the current research effort exhibit larger errors than in other studies~\cite{Drechsel:2007if, CLAS:2009ces} even though more data and reactions are analyzed.

{\it Methodology}--We calculate the TFFs of twelve selected $N^*$ and $\Delta$ states at the poles, based on the latest coupled-channel analyses~\cite{Mai:2023cbp} of $\pi N$, $\eta N$, and $K\Lambda$ electroproduction off the proton within the J{\"u}lich-Bonn-Washington (JBW) framework~\cite{Mai:2021vsw,Mai:2021aui,Mai:2023cbp}.  The input at the photon point and for the hadronic final-state interaction is provided by the J{\"u}lich-Bonn dynamical coupled-channel model. This unitary framework is based on the meson exchange picture and has been developed over the last decades, see Refs.~\cite{Ronchen:2012eg,Ronchen:2022hqk,Wang:2022osj} and references therein. A potential, which is derived from a chiral Lagrangian, is iterated  in a Lippmann-Schwinger-like equation to obtain the scattering amplitude.
The analytic properties of the approach, a prerequisite for determining resonances as poles on the unphysical Riemann sheet, have been discussed in Ref.~\cite{Doring:2009yv}, see also Ref.~\cite{Sadasivan:2021emk}. In contrast to the purely hadronic interactions, the interaction of a (real or virtual) photon with the meson-baryon states is parametrized by energy-dependent polynomials~\cite{Ronchen:2014cna} and the $Q^2$ dependence is introduced in a separable form that fulfills Siegert's condition~\cite{Mai:2021vsw} which is a consequence of gauge invariance. For a very brief summary of the formalism, see the Supplemental Material~\cite{SM}\footnote{The Supplemental Material includes Refs.~\cite{schweber1964,sterman1993,Zhang:2021hcl} that are relevant to the ``Time-Ordered Perturbation Theory'' (TOPT), based on which the Lippmann-Schwinger-like equation of this model has been constructed. }.

\begin{figure}[t]
\includegraphics[width=1\linewidth,trim=0.5cm 0 0.5cm 0]{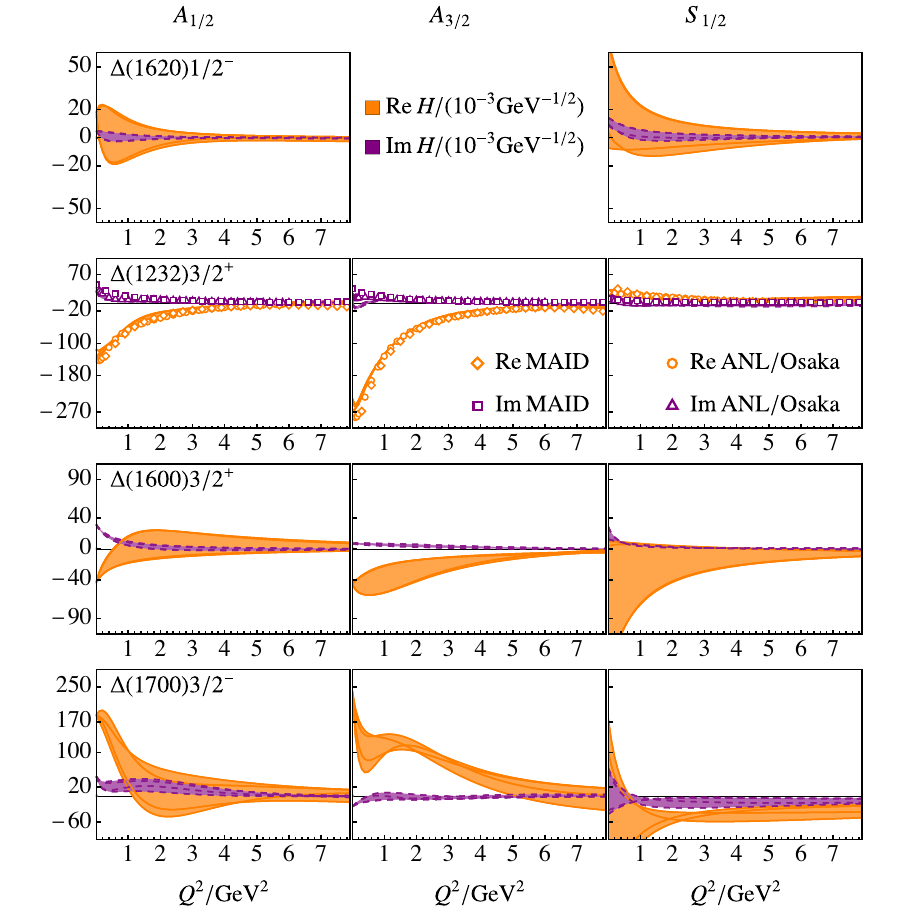}
	 \caption{$\Delta$ transition form factors of this work (colored bands). The orange (purple) band with solid (dashed) lines exhibits the real (imaginary) part. The bands represent the uncertainties of the extraction, see text for further explanation. All available literature results from MAID~\cite{Tiator:2016btt} and ANL-Osaka solutions are depicted by empty symbols~\cite{Kamano:2018sfb}. Note that there are no literature results (defined at the pole) for the states other than $\Delta(1232)$. 
  }
 \label{fig:TFF_allD} 
\end{figure}

\begin{figure}[t]
\includegraphics[width=1\linewidth,trim=0.5cm 0 0.5cm 0]{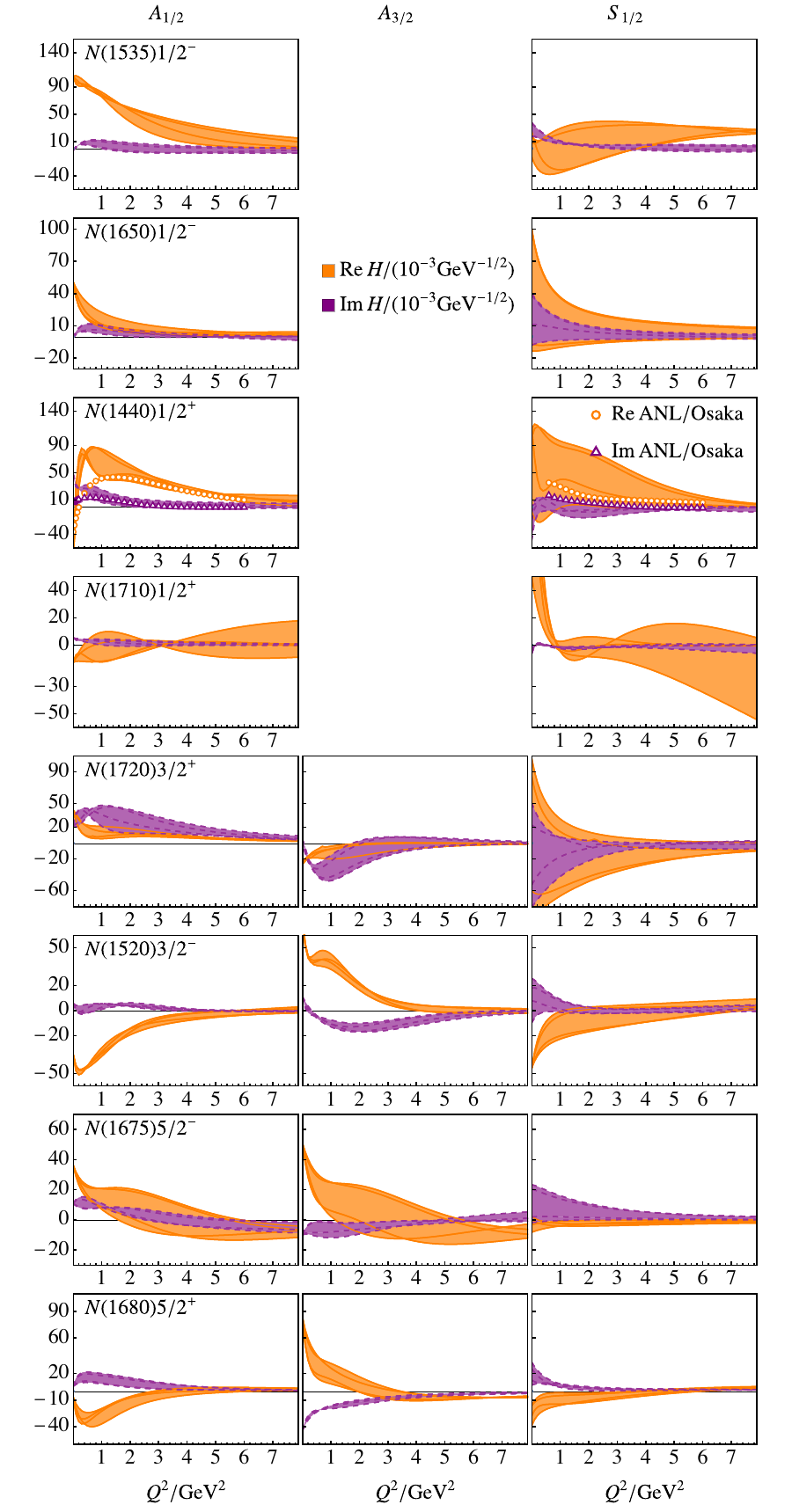}
\caption{$N^*$ transition form factors of this work (colored bands). See the caption of Fig.~\ref{fig:TFF_allD} for the explanations of the symbols. Note that there are no literature results (defined at the pole) for the states other than $N(1440)$. 
}
\label{fig:TFF_allN}
\end{figure}

The parameters of the resonance poles that are discussed in this work were determined in the ``J{\"u}Bo2017'' analysis~\cite{Ronchen:2018ury}, a simultaneous analysis of the channels $\pi N\to\pi N$, $\eta N$, $K\Lambda$, $K\Sigma$ and $\gamma p\to\pi N$, $\eta N$, $K\Lambda$. The corresponding electroproduction amplitudes were obtained in Ref.~\cite{Mai:2023cbp}, the first-ever coupled-channel analysis of pion-, eta- and $K\Lambda$ electroproduction off the proton, for $W\in[1.13,1.8]\,\GeV$ and $Q^2\in [0,8]\,\GeV^2$.

To define the TFFs, we may focus on the reactions $\gamma^* N\to \pi N$ and $\pi N\to \pi N$ because TFFs at resonance poles are reaction independent. Elastic pion-nucleon scattering is parametrized with the dimensionless partial waves $\tau^{l\pm}$ indicating orbital angular momentum $l$ and total spin $J=l\pm \nicefrac{1}{2}$~\cite{Ronchen:2012eg}. Pion electroproduction is parametrized via multipoles with definite $\gamma^* N$ helicity $h=\frac{1}{2},\frac{3}{2}$, isospin $I$, orbital angular momentum $l$, and total angular momentum $J$, denoted as  $\mathcal{A}_{h}^{l\pm,I}$, and $\mathcal{S}_{1/2}^{l\pm,I}$, the so-called helicity amplitudes. Note that the latter corresponds to the longitudinal polarization part, which do not contribute to any observables with $Q^2 = 0$.

The Laurent expansion at the resonance pole $z=z_p$ reads 
\begin{equation}
  \mathcal{H}_{h}^{l\pm,I}=\frac{\widetilde{\mathcal{H}}_{h}^{l\pm,I}}{z-z_p}+\cdots\ ,\quad
	\tau^{l\pm,I}=\frac{\widetilde{R}^{l\pm,I}}{z-z_p}+\cdots\,,
\end{equation}
where $\mathcal{H}$ denotes either $\mathcal{A}$ or $\mathcal{S}$, and $\widetilde{\mathcal{H}}$, $\widetilde{R}$ denote residues. Finally, following Ref.~\cite{Workman:2013rca}, the transition form factors $H_h^{l\pm,I}$ (i.e., $A_{\nicefrac{1}{2}},\,A_{\nicefrac{3}{2}},\,S_{\nicefrac{1}{2}}$), which depict the $Q^2$ dependence of the residues of the helicity amplitudes, are reaction independently defined as: 
\begin{equation}
\label{TFFdef}
	H_{h}^{l\pm,I}(Q^2)=
	C_I\sqrt{\frac{p_{\pi N}}{\omega_0}\frac{2\pi(2J+1)z_p}{m_N\widetilde{R}^{l\pm,I}}}\widetilde{\mathcal{H}}_{h}^{l\pm,I}(Q^2)\,,
\end{equation}
with $\omega_0$ the energy of the photon at $Q^2=0$, $m_N$ the nucleon mass, and the isospin factor~\cite{Drechsel:1998hk} $C_{1/2}=-\sqrt{3}$ and $C_{3/2}=\sqrt{2/3}$. In the remaining text, the superscripts $l\pm,I$ are suppressed, since they are known for each resonance. 

{\it Results}--Coming to the results, twenty-six states are found in Ref.~\cite{Ronchen:2018ury}, but in this work only the twelve with pole mass $\text{Re}(z_p)\leq 1.8\,\GeV$ and orbital angular momentum $l\leq 3$ are chosen, corresponding to the energy range and the truncation of the angular momentum in Ref.~\cite{Mai:2023cbp}, namely $N(1535)$, $N(1650)$, $N(1440)$, $N(1710)$, $N(1720)$, $N(1520)$, $N(1675)$, $N(1680)$, $\Delta(1620)$, $\Delta(1232)$, $\Delta(1600)$, $\Delta(1700)$. The TFFs for all these states are shown in Figs.~\ref{fig:TFF_allD} and \ref{fig:TFF_allN} and will be available in numerical form~\cite{JBW-homepage}. The amplified figures for $\Delta(1232)$ and $N(1440)$ can be found in the Supplemental Material~\cite{SM}. 

There are four fits in Ref.~\cite{Mai:2023cbp} with similar $\chi^2$ which roughly give the scale of uncertainties associated with the extraction; as mentioned before, these uncertainties are larger than those of other studies owing to an extensive exploration of the parameter space. There is a more detailed discussion in the Supplemental Material~\cite{SM}. 

As for $\Delta(1232)$, see Fig.~\ref{fig:TFF_allD}, the four fits of Ref.~\cite{Mai:2023cbp} do not deviate much from one another, reflected by a rather narrow band of the transition form factors. Putting this into perspective with the available literature values, we
show in the same figure the preliminary results of the ANL-Osaka model from an updated version~\cite{Kamano} of Ref.~\cite{Kamano:2018sfb}, as well as results of the MAID model~\cite{Tiator:2016btt}. The real parts are in good agreement with our global estimations within the uncertainties due to different solutions.
Some discrepancies can be seen in the curves of $\text{Im}\,A_{3/2}$ in the small-$Q^2$ region, which are due to the difference at the photon-point ($Q^2=0\,\GeV^2$). Our result is actually closer to an older version of ANL-Osaka model in Ref.~\cite{Julia-Diaz:2006ios}, as plotted in Fig.~7 of Ref.~\cite{Kamano:2018sfb}.

As for the TFFs of the low-lying resonances determined in Ref.~\cite{Mokeev:2023zhq}, including the $\Delta(1600)$, our results are not in disagreement, given their slightly different definitions from Breit-Wigner resonances, although the uncertainties found here are larger. We note in passing that the Breit-Wigner parametrization can only be employed safely for a few resonances, as the basis assumption of a slowly varying background is rarely fulfilled due to many open channels.

\begin{figure}[t]
\includegraphics[width=0.9\linewidth,trim=0 0 0 0]{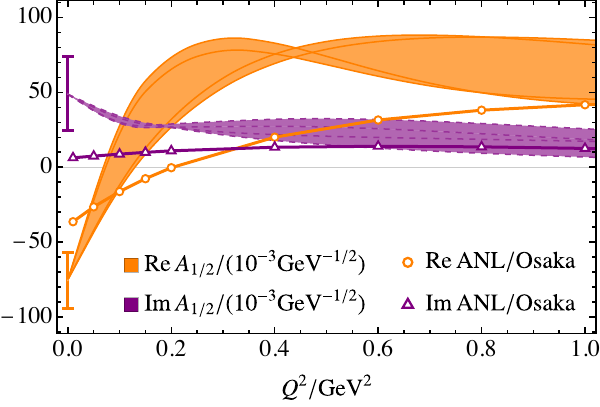}
\caption{$N(1440)$ transition form factors at small $Q^2$ of this work (colored bands) in comparison with the ANL-Osaka solution. The error bars at $Q^2=0\,\GeV^2$ depict the uncertainties of the photoproduction solution at the pole from Ref.~\cite{Ronchen:2018ury}. See the caption of Fig.~\ref{fig:TFF_allD} for the explanations of the other symbols.
}
\label{fig:N1440zoom}
\end{figure}

As for $N(1440)$, see Figs.~\ref{fig:TFF_allN} and ~\ref{fig:N1440zoom}, in all four fits we consistently obtain a zero crossing in $\text{Re\,}A_{1/2}$, which is also found in Refs.~\cite{Obukhovsky:2011sc,Wilson:2011aa,Segovia:2015hra,Aznauryan:2018okk,Obukhovsky:2019xrs,Burkert:2017djo,Chen:2018nsg,Julia-Diaz:2006ios,Kamano:2018sfb}. 
The zero occurs at $Q^2=(0.1\pm 0.04)$~GeV$^2$ which is smaller than in Breit-Wigner determinations, and also smaller than in the ANL-Osaka pole determination~\cite{Kamano:2018sfb} (zero at $Q^2=0.2$~GeV$^2$). Such a zero transition indicates that the core of $N(1440)$ can be explained as a radial excitation of the nucleon~\cite{Ramalho:2023hqd}. However, the structure of the $N(1440)$ is a complicated problem: the contributions from the meson clouds or two-hadron systems seem not to be negligible~\cite{Meissner:1984un,Zahed:1984qv,Obukhovsky:2011sc,Segovia:2015hra}; in some models including the present one, the three-quark core is even less important than the other components~\cite{Sekihara:2021eah,Wang:2023snv}. Especially, the $N(1440)$ is always dynamically generated from the $t$- and $u$- channel hadron-exchange potentials, without the need for a genuine $s$-channel state~\cite{Krehl:1999km}. This is verified by Ref.~\cite{Wang:2023snv}. Note that Refs.~\cite{Li:1991sh,Li:1991yba} have pointed out the $S_{1/2}$ of a hybrid $N(1440)$ state should be zero. We do not see strong indications of a vanishing $S_{1/2}$: the ${\rm Re}\,S_{1/2}$ in fit 1 is smaller but still non-zero. 

\begin{figure}[b]
\includegraphics[width=0.9\linewidth,trim=0.5cm 0 0.5cm 0]{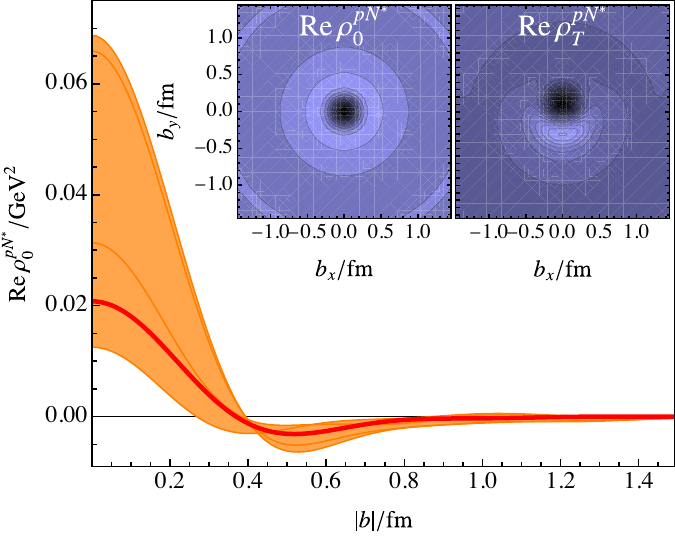}
	 \caption{Transverse charge density (unpolarized $\rho_0^{pN^*}$, polarized along x-axis $\rho_T^{pN^*}$) of the $p\to N(1440)$ transition as a function of the transverse position $b$ in the $xy$-plane. The orange band (thick red line) depicts the uncertainty band of our determination (the result using the MAID 2007 helicity couplings~\cite{Drechsel:2007if}). The inset shows corresponding coordinate decompositions with light/dark shades representing negative/positive values.}
 \label{fig:ChargeDensityRoper} 
\end{figure}

The four Roper helicity amplitudes determined in this study allow one to determine the transverse charge density $\rho$ of the resonance transitions~\cite{Tiator:2009mt,Tiator:2008kd,Ramalho:2023hqd} that provide additional insights into the resonance structure. Results are shown in Fig.~\ref{fig:ChargeDensityRoper}. We use the complex Roper pole position and complex helicity amplitudes for the calculation of the Dirac and Pauli form factors~\cite{Ramalho:2023hqd}. We then only show the real part of $\rho$ in the figure to have a point of comparison with other determinations.

The radially symmetric unpolarized case, $\rho_0^{pN^*}$, is shown in the main plot and left inset. 
As in Ref.~\cite{Tiator:2009mt} we observe a positively charge center surrounded by a weakly negatively charged region. In fact, the MAID 2007 solution (red line) induces a charge distribution well within the uncertainties of our solutions (orange band), despite the different zero transitions of Re~$A_{1/2}$ in MAID 2007~\cite{Drechsel:2007if} and the present solution.

The right inset shows the case $\rho_T^{pN^*}$ of a transversely polarized $p$ and $N^*$ along the $x$-axis for one of the four solutions that lies towards the center of the uncertainty band. 
In this case, radial symmetry is no longer given and an electric dipole moment is induced.
 
{\it Conclusion}--To summarize, this letter reports the baryon transition form factors for $N^*$ and $\Delta$ states defined at the resonance poles. For the first time, this is achieved by analyzing more than one final state at the same time, for altogether $\sim 10^5$ data. Additional constraints are provided by a large amount of pion- and (real) photon-induced single-meson data. Special emphasis is paid to a more extensive exploration of the parameter space of the analysis framework, resulting in realistic uncertainties for the TFFs reflecting the data situation and inherent ambiguities.

Twelve selected states are studied based on the multipoles determined in Ref.~\cite{Mai:2023cbp}. For the $\Delta(1232)$ and $N(1440)$ states, this work gives compatible results with  other studies. For many other states, the present determination is the first of its kind at resonance poles, providing truly reaction-independent TFFs; qualitative comparisons with Breit-Wigner determinations of other studies show no obvious disagreement.

Future data from JLab and other facilities will allow to extend the present analysis to higher photon virtualities mapping the transition region to perturbative QCD. Such data would also allow for the exploration of higher-lying resonances that could be of hybrid nature manifested in an unusual $Q^2$ dependence of their transition form factors~\cite{JLABhybrid}. In turn, this study also can provide a guidance for future experiments, pinpointing the kinematic regions which are most relevant and least determined to reduce the error bars on the TFF's. 

~\\

{\it Acknowledgements}--We thank Hiroyuki Kamano, Toru Sato, Craig Roberts, Christian Fischer and Lothar Tiator for helpful discussions. The authors gratefully acknowledge computing time on the supercomputer JURECA~\cite{JURECA} at Forschungszentrum J\"ulich under grant no. ``baryonspectro". This work is supported by the NSFC and the Deutsche Forschungsgemeinschaft (DFG, German Research Foundation) through the funds provided to the Sino-German Collaborative Research Center TRR110 “Symmetries and the Emergence of Structure in QCD” (NSFC Grant No. 12070131001, DFG Project-ID 196253076-TRR 110). Further support by the CAS through a President’s International Fellowship Initiative (PIFI) (Grant No. 2024PD0022) is acknowledged. This work is also supported by the MKW NRW under the funding code NW21-024-A and by the Deutsche Forschungsgemeinschaft (DFG, German Research Foundation) – 491111487. The work of RW and MD is supported by the U.S. Department of Energy grant DE-SC0016582 and Office of Science, Office of Nuclear Physics under contract DE-AC05- 06OR23177 (MD). This work contributes to the aims of the U.S. Department of Energy ExoHad Topical Collaboration, contract DE-SC0023598. The work of TM is supported by the PUTI Q1 Grant from University of Indonesia under contract 
No. NKB-441/UN2.RST/HKP.05.00/2024. The work of UGM and DR is further supported  by the Deutsche Forschungsgemeinschaft (DFG,
German Research Foundation) as part of the CRC 1639 NuMeriQS – project
no. 511713970.
\bibliography{TFF_Paper}
\clearpage
\end{document}


\title{Global Data-Driven Determination of Baryon Transition Form Factors: \\Supplemental Material}

\begin{abstract}
This note is the supplemental material to the letter, which briefly summarizes the formalism of the coupled-channel dynamics of the model employed in this work, and shows the amplified figures for the $N(1440)$ and $\Delta(1232)$ states. The uncertainties of this work are also discussed here in a slightly more detailed manner. 
\end{abstract}
\maketitle

\section{Coupled-Channel Formalism}\label{app:dy}
The master equation describing the hadronic dynamics in this model is the following Lippmann-Schwinger-like equation: 
\begin{equation}\label{scequ}
\begin{split}
  &T_{\mu\nu}(p'',p',z)=V_{\mu\nu}(p'',p',z)\\
  &+\sum_{\kappa}\int_0^\infty p^2 dp V_{\mu\kappa}(p'',p,z)G_{\kappa}(p,z)T_{\kappa\nu}(p,p',z)\ ,
\end{split}
\end{equation}
where $T$ is the hadronic scattering amplitude, $V$ is the interaction potential, $p'$
and $p''$ are the three-momenta of the initial and final states in the center-of-mass frame, respectively,
$z$ is the center-of-mass energy. Further, $\mu,\nu$ and $\kappa$ are the channel labels
denoting the meson-baryon system with specific quantum numbers. In this work the hadronic channel space includes 
$\pi N$, $\pi\pi N$, $\eta N$, $K\Lambda$, and $K\Sigma$ (though $K\Sigma$ photo- and electroproduction is not included). 
The three-body channel $\pi\pi N$ is simulated by three effective channels 
$\pi\Delta$, $\sigma N$, and $\rho N$. Moreover, $G_{\kappa}(p,z)$ is the propagator of the intermediate channel: 
\begin{equation}\label{Gdef}
	G_{\kappa}(z,p)=\left[z-E_\kappa-\omega_\kappa-\Sigma_\kappa(z,p)+i0^+\right]^{-1}\ ,
\end{equation}
with $E_\kappa,\,\omega_\kappa$ being the energies of the baryon and the meson in channel $\kappa$,
respectively, with a relativistic dispersion relation, e.g. $E_\kappa=\sqrt{p^2+M_\kappa^2}$.
For the effective channels, $\Sigma_\kappa$ is the self-energy function of the unstable
particle ($\Delta,\sigma,\rho$), otherwise it is zero. 

The interaction potentials, as well as the propagators, are derived from (old-fashioned)
time-ordered perturbation theory (TOPT)~\cite{schweber1964} rather than the modern covariant convention.
In combination with a partial-wave projection, the application of TOPT reduces the four-dimensional 
Bethe-Salpeter equation to the one-dimensional integral equation. This is equivalent to a certain kind of reduction 
of the full relativistic Bethe-Salpeter equation, with the zeroth component $p^0$ of the intermediate momentum 
integrated out and the residues from the negative frequencies discarded~\cite{sterman1993}. Note that the missing 
terms from the negative frequencies does not have significant impacts on the results, see e.g. Ref.~\cite{Zhang:2021hcl}. For 
the details of the interaction potentials, see Ref.~\cite{Ronchen:2012eg}. 

The multipoles are obtained from the hadronic scattering amplitude through the following equation satisfying Watson's 
final state theorem: 
\begin{equation}\label{scequgam}
\mathcal{M}_{\mu\gamma}=V_{\mu\gamma}
+\sum_{\kappa}\int_0^\infty p^2 dp T_{\mu\kappa}G_{\kappa}V_{\kappa\gamma}\ ,
\end{equation}
with $\gamma$ denoting the $\gamma N$ channel. The potential $V_{\mu\gamma}$ is phenomenologically formulated, for details see Ref.~\cite{Ronchen:2014cna}. Furthermore, the non-trivial part depicting the electroproduction is $\mathcal{M}_{\mu\gamma^*}$, with $\gamma^*$ denoting the system of the virtual photon and the nucleon. It can be obtained by adding the $Q^2$ dependence to the potential, through modifications of the $V_{\mu\gamma}$ considering all additional requirements such as Siegert's theorem. The details can be found in Refs.~\cite{Mai:2021vsw,Mai:2021aui,Mai:2023cbp}. 

\section{Amplified figures for two states}\label{app:figs}
\begin{figure*}[bt]
	\centering
	\includegraphics[width=1.0\textwidth]{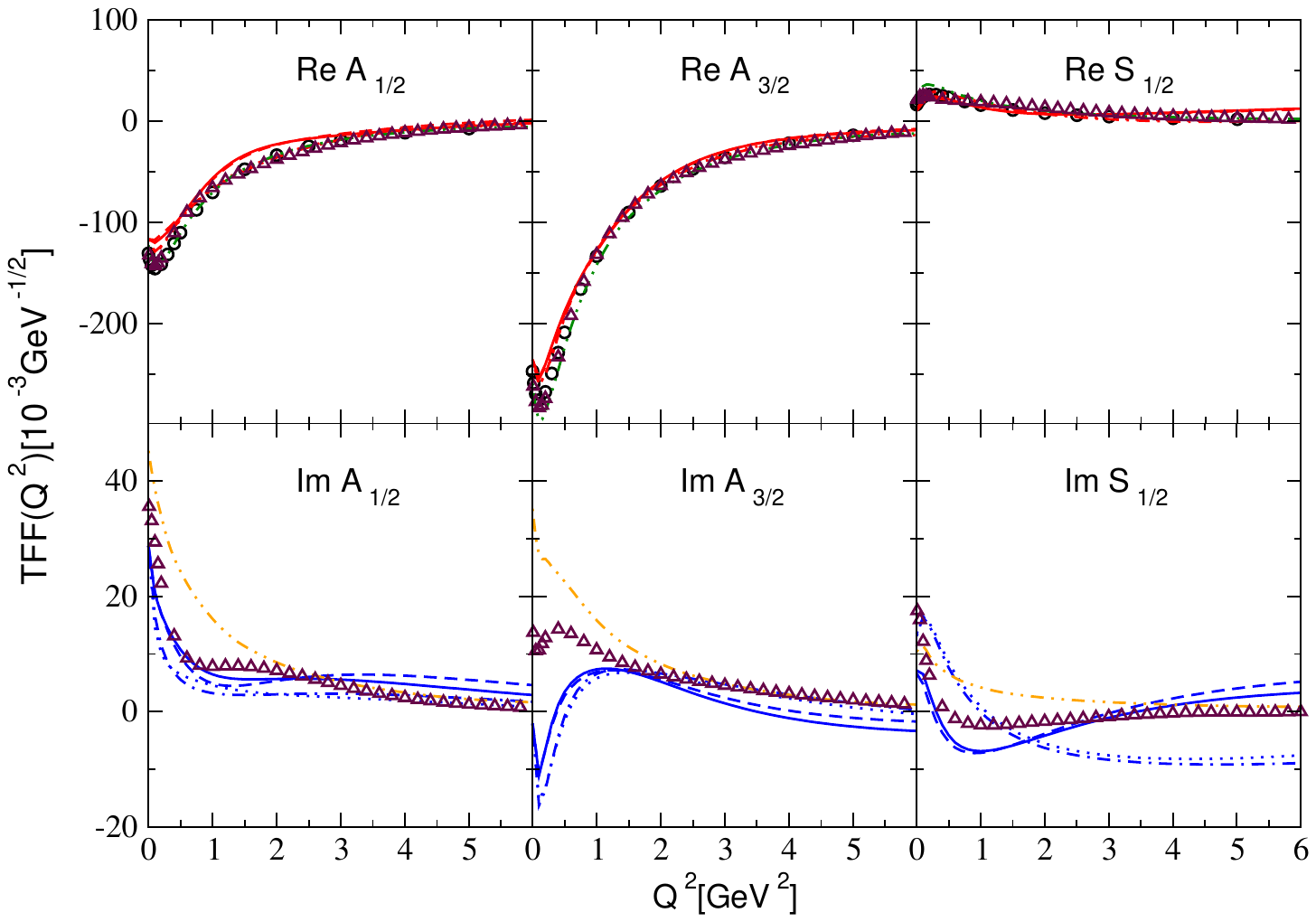}
	\caption{The transition form factors of $\Delta(1232)$. Solid, dashed, dotted, dash-dotted lines: results of this work. Double-dotted lines: results from Ref.~\cite{Tiator:2016btt} based on the MAID results. Black circles: original MAID results (real-valued) employed in Ref.~\cite{Tiator:2016btt}. Triangles: preliminary results of the ANL-Osaka model~\cite{Kamano,Kamano:2018sfb}. }
	\label{fig:1232}
\end{figure*}

\begin{figure*}[bt]
	\centering
	\includegraphics[width=1.0\textwidth]{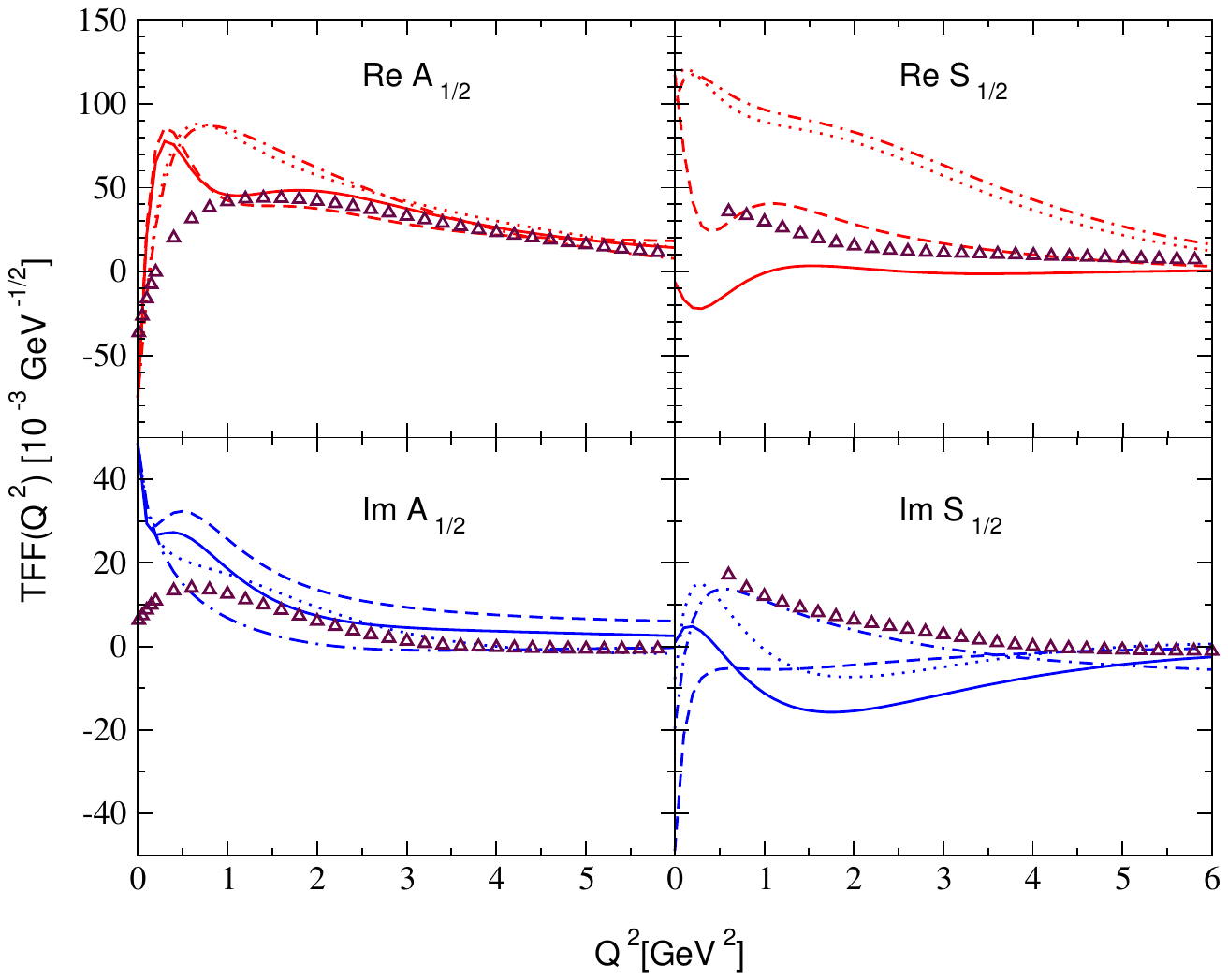}
	\caption{The transition form factors of $N^*(1440)$. Notations are the same as Fig.~\ref{fig:1232}. }
	\label{fig:1440}
\end{figure*}

Here we show the amplified figures of the transition form factors of $\Delta(1232)$ and $N(1440)$. The form factors of the $\Delta(1232)$ are compared to the preliminary results of the ANL-Osaka model (an updated version~\cite{Kamano} of Ref.~\cite{Kamano:2018sfb}), as well as the MAID results extracted in Ref.~\cite{Tiator:2016btt}, see Fig.~\ref{fig:1232}. The results of $N(1440)$ are compared to Refs.~\cite{Kamano,Kamano:2018sfb} in Fig.~\ref{fig:1440}. 
\section{Discussions on the uncertainties}\label{app:err}
In this work, the uncertainties of the transition form factors stem from the differences among the four fit solutions in Ref.~\cite{Mai:2023cbp}. Within one model parametrization, the data has been analyzed with both unweighted and weighted $\chi^2$ (weights on the $K\Lambda$ data). In each case, two different local minima have been found, as the four solutions. We believe those four solutions estimate a full exploration of the parameter space. More details can be found in Sect.~3 of Ref.~\cite{Mai:2023cbp}. 

It has been mentioned in the letter that one source of the uncertainties is the incomplete data base of electroproduction. Specifically, the data required to eliminate ambiguities in Complete Experiments or Truncated Partial-Wave Analysis is a topic that continues to be studied in, for example, Refs.~\cite{Tiator:2017cde,JointPhysicsAnalysisCenter:2023gku}. In Ref.~\cite{Tiator:2017cde}, a rare case with sufficient data is provided by pion electroproduction data from Kelly~\cite{JeffersonLabHallA:2005wfu}. When inconsistent datasets are fitted (mainly cross sections with little polarization data), an exploration of the parameter space gives amplitude bands which can be compared to other fits. Clearly, the reliability will increase with more precise and numerous data. In Ref.~\cite{Mai:2023cbp}, focused on kaon electroproduction, we showed the influence of unfitted polarization data. In addition, the ``variations'' of the current database has also been shown, providing a picture of how the data in different regions of the center-of-mass energy $W$ and the photon virtuality $Q^2$ affect the uncertainties. See Fig~9 of Ref.~\cite{Mai:2023cbp}. 

\bibliographystyle{h-physrev}
\bibliography{TFF_SM}